\documentclass[aps,preprint,superscriptaddress]{revtex4-1}

\include{preamble}
\usepackage{amsmath,amsfonts,amsthm,amssymb}
\usepackage{graphicx}

\DeclareMathOperator\erfc{erfc}

\begin{document}
%lhead{}
%\chead{}
%\rhead{Working Paper}

%\preprint{Working Paper}

\title{Simple Formula for Marcus-Hush-Chidsey Kinetics}
% for Electron Transfer at Metal Electrodes}

\author{Yi Zeng}
\affiliation{Department of Mathematics, Massachusetts Institute of Technology, Cambridge, Massachusetts 02139, USA}

\author{Raymond Smith}
\affiliation{Department of Chemical Engineering, Massachusetts Institute of Technology, Cambridge, Massachusetts 02139, USA}

\author{Peng Bai}
\affiliation{Department of Chemical Engineering, Massachusetts Institute of Technology, Cambridge, Massachusetts 02139, USA}

\author{Martin Z. Bazant}
\thanks{Corresponding author: bazant@mit.edu}
\affiliation{Department of Mathematics, Massachusetts Institute of Technology, Cambridge, Massachusetts 02139, USA}
\affiliation{Department of Chemical Engineering, Massachusetts Institute of Technology, Cambridge, Massachusetts 02139, USA}

\date{\today}

% ---------------------------------------------------------------------------------
\begin{abstract}
The Marcus-Hush-Chidsey (MHC) model is well known in electro-analytical chemistry as a successful microscopic theory of outer-sphere electron transfer at metal electrodes, but it is unfamiliar and rarely used in electrochemical engineering.  One reason may be the difficulty of evaluating the MHC reaction rate, which is defined as an improper integral of the Marcus rate over the Fermi distribution of electron energies. Here, we report a simple analytical approximation of the MHC integral that interpolates between exact asymptotic limits for large overpotentials, as well as for large or small reorganization energies, and exhibits less than 5\% relative error for all reasonable parameter values.  This result enables the MHC model to be considered as a practical alternative to the ubiquitous Butler-Volmer equation for improved understanding and engineering of electrochemical systems.
\end{abstract}

%\keywords{}
\maketitle

% -----------------------------------------------------------------------------------------
% New Section
% -----------------------------------------------------------------------------------------

\section{Introduction}

The microscopic theory of electron transfer~\cite{kuznetsov_book,bard_book} has been developed and tested in electroanalytical chemistry for almost seventy years since the pioneering work of Marcus~\cite{marcus1956,marcus1964arpc,marcus1993}. Although much of the early work focused on homogeneous electron transfer in solution, the theory was also extended to heterogeneous electron transfer at electrodes~\cite{marcus1993,marcus1965,henstridge2012} and found to accurately predict Faradaic reaction kinetics for both liquid~\cite{chidsey1991,migliore2011,henstridge2012} and, more recently, solid~\cite{bai2014} electrolytes.  For metal electrodes, however, the theory is complicated by the need to integrate the Marcus rate over the Fermi-Dirac distribution of electrons.  This integral cannot be evaluated in closed form in terms of elementary functions and has only been approximated (in certain limits) by relatively cumbersome series expansions~\cite{oldham2011,migliore2011,migliore2012}.

Partly for this reason, despite its successes, the theory is  rarely used and poorly known in engineering.  Instead, standard mathematical models are based on the phenomenological Butler-Volmer (BV) equation~\cite{bockris_book,newman_book}, which has the appeal of a simple analytical formula that fits many experimental measurements, even though it lacks a clear physical basis. The goal of this work is to derive an equally simple formula for the microscopic theory.

% -----------------------------------------------------------------------------------------
% New Section
% -----------------------------------------------------------------------------------------

\section{ Background }

For the simple redox reaction, R $\leftrightarrow$ O + e$^-$, the BV reductive and oxidative reaction rates, are expressed as,
\begin{equation}
\begin{aligned}
    &k_{red}^{BV}(\eta,\alpha)= k_0^{BV}\exp\left(\frac{-\alpha e\eta}{k_BT}\right),
    \\
    &k_{ox}^{BV}(\eta,\alpha) = k_0^{BV}\exp\left(\frac{(1-\alpha)e\eta}{k_BT}\right),
    \label{}
\end{aligned}
\end{equation}
where $k_0^{BV}$ is the rate constant, $\alpha$ the charge transfer coefficient, $e$ the elementary charge, $\eta$ the applied overpotential, $k_B$ Boltzmann's constant and T the temperature.  The net reduction current is proportional to the difference in forward and backward rates, $I \propto k_{red} - k_{ox}$, in the standard form of the BV equation.    The ratio of forward and backward rates satisfies the de Donder relation,
\begin{equation}
\frac{k_{red}}{k_{ox}} =\exp\left(-\frac{e\eta}{k_BT}\right)    \label{eq:dedonder}
\end{equation}
which is a general constraint from statistical thermodynamics for thermally activated chemical kinetics~\cite{bazant2013,sekimoto_book}. The BV model asserts that the reaction rate in either direction follows the Tafel relationship, in which the thermodynamic driving force is a constant fraction of the applied overpotential.   This dependence is empirical but can be justified by various phenomenological models~\cite{bockris_book,bard_book}, where the electrostatic energy of the (ill-defined) transition state of the reaction is an average of that in the reduced and oxidized states, weighted by the charge transfer coefficient~\cite{bazant2013}.

In contrast, the microscopic theory of outer-sphere electron transfer focuses on solvent reorganization prior to iso-energetic electron transfer~\cite{marcus1993,kuznetsov_book,bard_book}. In the simplest form of the theory, the free energy of the reduced and oxidized states has the same harmonic dependence on a reaction coordinate for solvent reorganization (such as local dielectric constant of the solvation shell), before and after electron transfer. For the same redox reaction above, the reaction rates take the form~\cite{sutin1983,chidsey1991,bazant2013},
\begin{equation}
    k_{red/ox}^{M}(\Delta G) 
    = k_0^{M}\exp\left(-\frac{(\Delta G \pm \lambda)^2}{4\lambda k_BT}\right),
    \label{}
\end{equation}
where $\Delta G$ is the free energy change upon reduction,
and $\lambda$ is the reorganization energy,  i.e. the free energy required
to completely reorganize the local atomic configuration of one state to the other
state \emph{without} charge transfer.

If the redox reaction occurs at an electrode, electrons in the metal electrode occupying different energy levels around the Fermi level may all participate in the reaction, which results in multiple intersections between two families of parabolae~\cite{marcus1965}. Although this principle was first identified decades ago, the importance of incorporating the Fermi-Dirac distribution of electrons/holes into the classical Marcus theory was not widely recognized until Chidsey found perfect agreement between the modified rate equation and the curved Tafel plot obtained from his seminal experiments on redox active self-assembled monolayers (SAMs)~\cite{chidsey1991}. The rate equation implemented by Chidsey, now known as the Marcus-Hush-Chidsey (MHC)~\cite{henstridge2012} or Marcus-DOS model~\cite{finklea2001}, can be written as,
\begin{equation}
    k_{ox/red}^{MHC}(\eta) = A\int_{-\infty}^{\infty}
    \exp\left(-\frac{(x -\lambda \pm e\eta)^2}{4\lambda k_BT}\right)
    \frac{dx}{1 + \exp(x/k_BT)},
    \label{eqn:MHC_full}
\end{equation}
where $A$ is the pre-exponential factor, accounting for the electronic coupling strength and the electronic density of states (DOS) of the
electrode. The first term in the integrand is the classical Marcus rate for the transfer of an electron of energy $x$ relative to the Fermi level, and the second factor is the Fermi-Dirac distribution assuming a uniform DOS. The reductive and oxidative reaction rates satisfy the de Donder relationship, Eq. ~\ref{eq:dedonder}, as well as a ``reciprocity relationship'' noted by Oldham and Myland~\cite{oldham2011}, $k_{ox}^{MHC}(-\eta) = k_{red}^{MHC}(\eta)$.

\begin{figure}
\centering
 \includegraphics[width=2.8in]{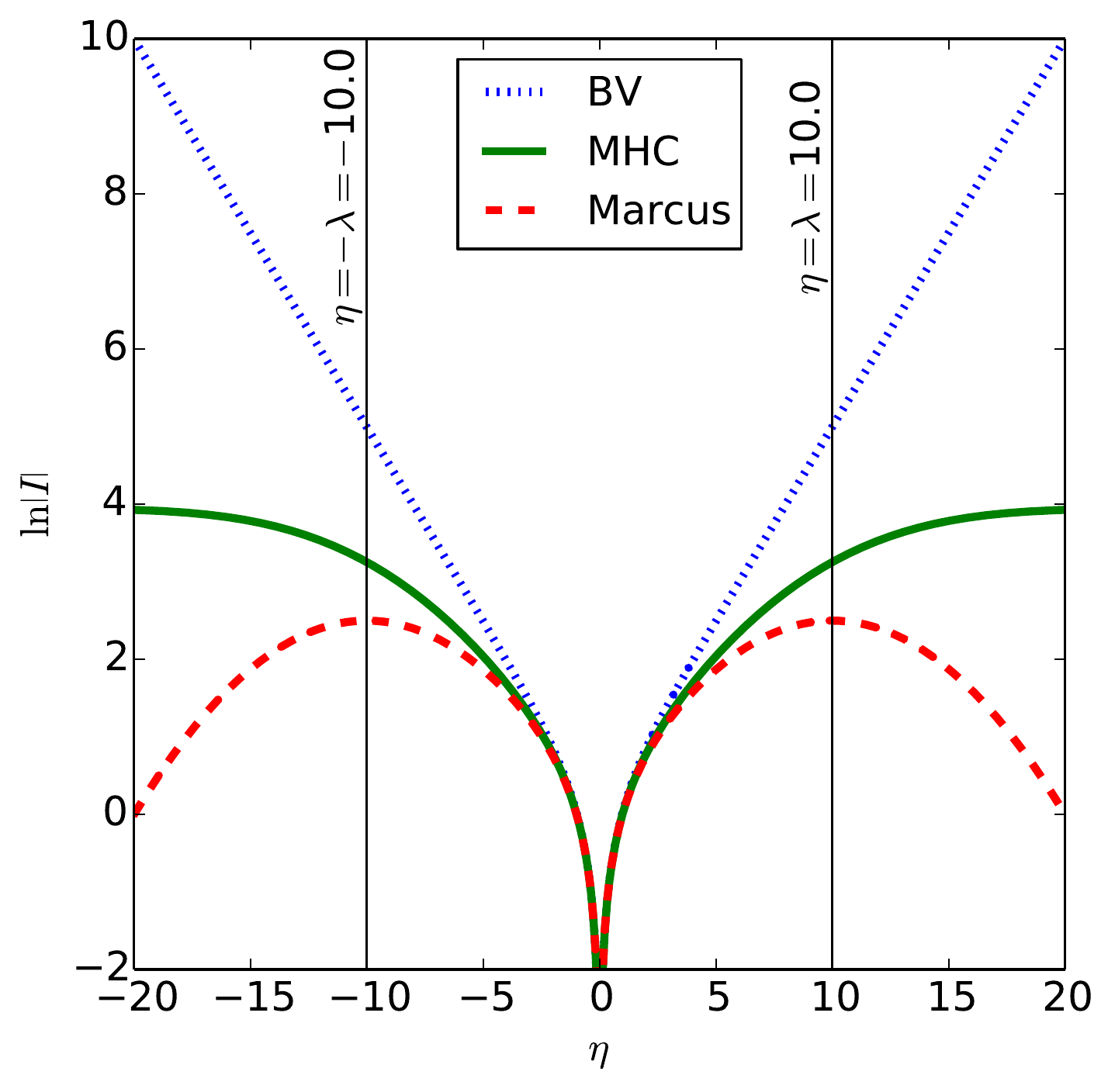}
\caption{ Dimensionless Tafel plots of Butler-Volmer kinetics (BV) with charge transfer coefficient $\alpha=\frac{1}{2}$ compared with Marcus (M) and Marcus-Hush-Chidsey (MHC) kinetics with reorganization energy $\lambda$ (scaled to  the thermal energy $k_BT$). The absolute value of the current $|I|$ scaled to the exchange current $I_0$ is plotted on a logarithmic scale versus the overpotential $\eta$ scaled to the thermal voltage, $k_BT/e$. The M and MHC curves assume a typical value~\cite{chidsey1991,bai2014} of the dimensionless reorganization energy, $\lambda=10$ scaled to $k_BT$.   \label{fig:Tafel} }
\end{figure}

The three models are compared on a Tafel plot in Figure ~\ref{fig:Tafel}, which highlights dramatic differences in the predicted rate for large overpotentials.  While the BV rate increases exponentially without bound along a traditional ``Tafel line", the Marcus rate reaches a maximum at the reorganization voltage ($\eta = \lambda/e$) and then decreases rapidly (as a Gaussian) along an inverted parabola.  The latter is the famous ``inverted region" predicted by Marcus for homogeneous electron transfer~\cite{marcus1993}.  The MHC model predicts a curved Tafel plot that neither diverges nor decays, but instead approaches a constant reaction-limited current.

\begin{figure}
    \centering
    (a)
    \includegraphics[width=2.8in]{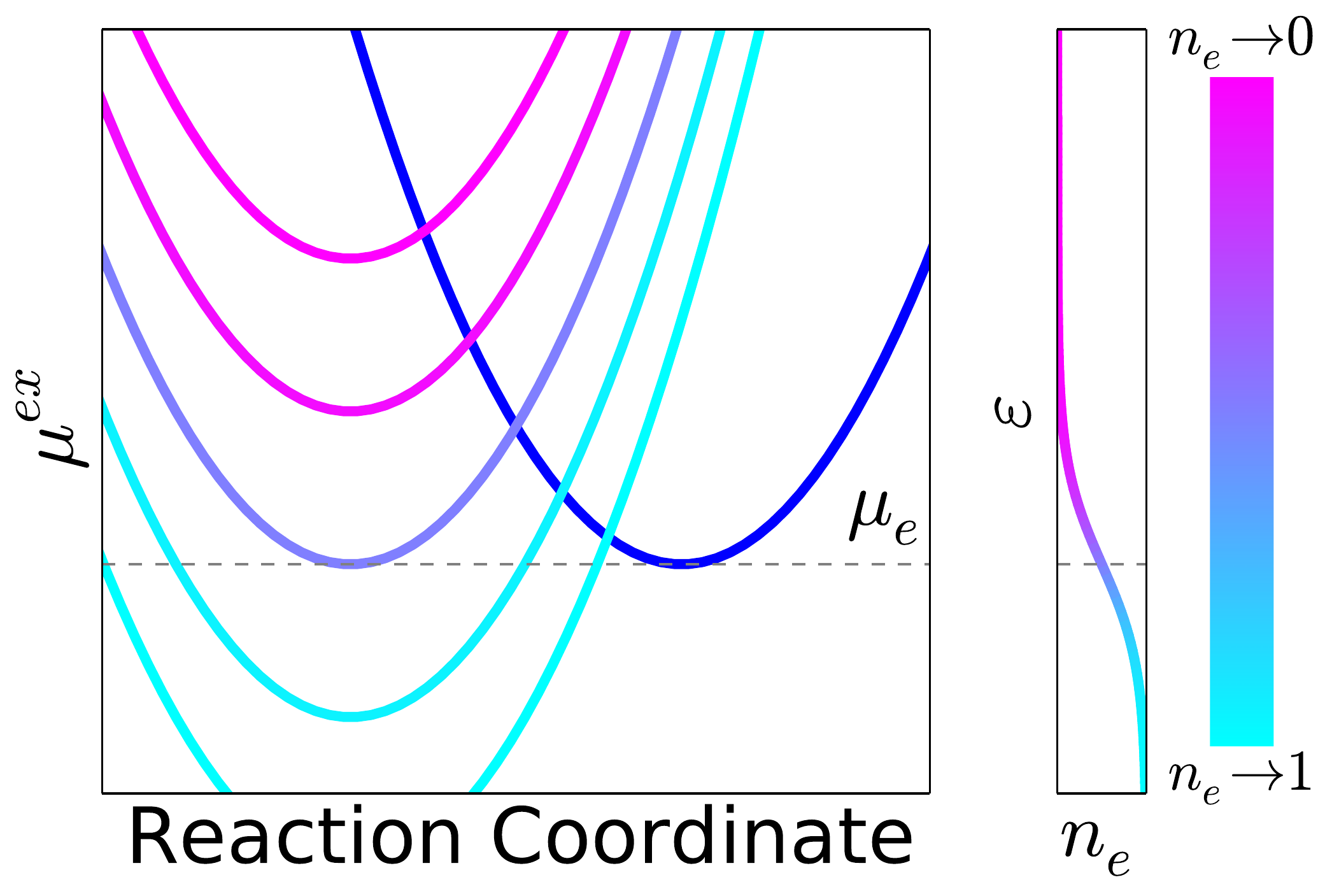}
    \\
    (b)
    \includegraphics[width=2.8in]{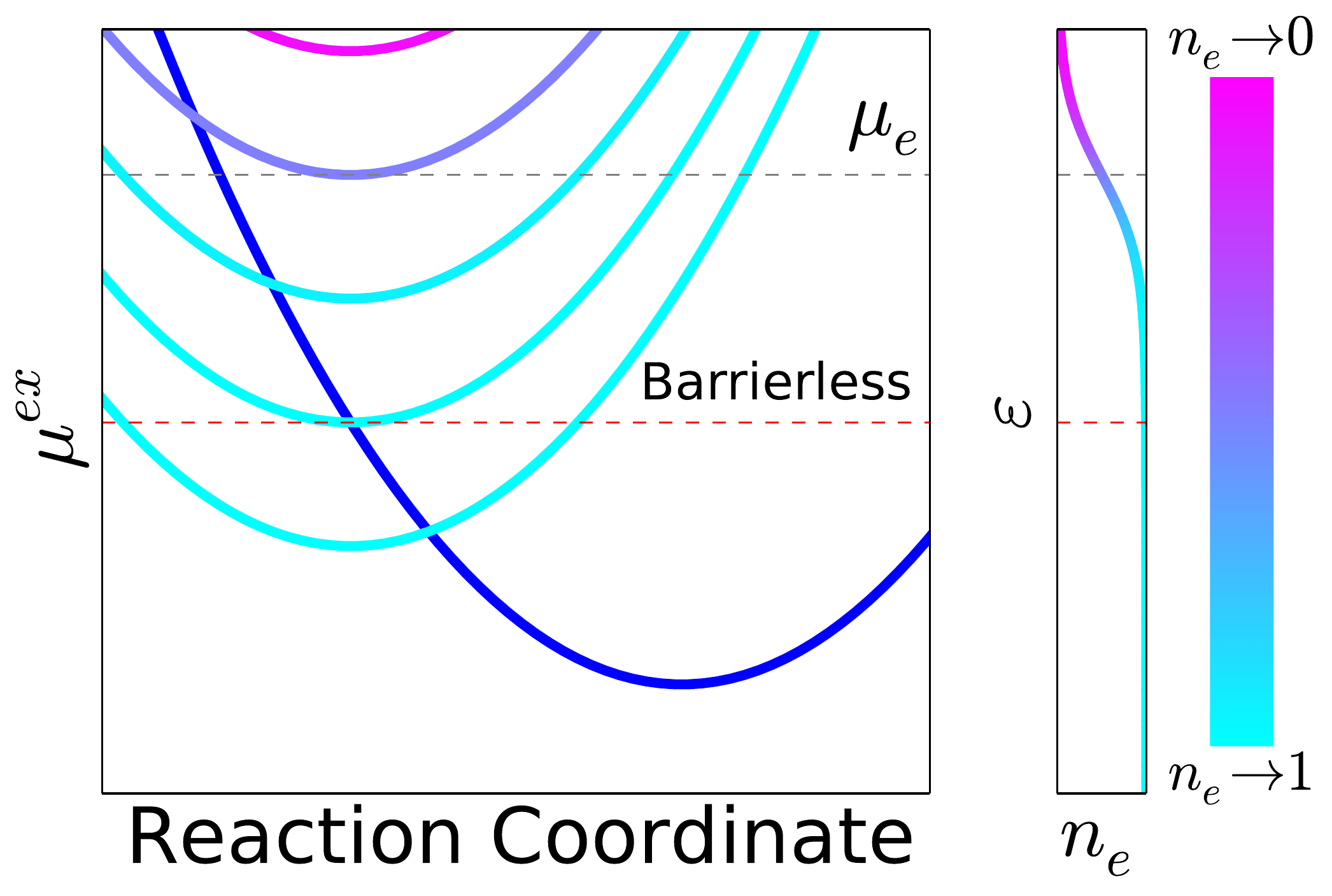}
    \caption{ Physical interpretation of MHC kinetics for the Faradaic reaction, O + $e^{-} \rightarrow$ R, at a metal electrode. In each panel, a parabola for the free energy (or more precisely, excess electrochemical potential~\cite{bazant2013}) of the reduced state (R, right) versus reorganization reaction coordinate intersects families of parabolae for the free energy of the oxidized state plus the free electron (O + e$^-$, left), sampled from the Fermi-Dirac distribution with electron energies, $\varepsilon$, shown. (a) Exchange process at zero overpotential, dominated by electrons near the Fermi level following Marcus kinetics. (b) Reaction-limited current at large negative overpotential, dominated by lower-energy electrons below the Fermi level undergoing barrier-less transitions.          \label{fig:MHC_curves} }
%    \caption{ Physical interpretation of MHC kinetics for the Faradaic reaction, O + $e^{-} \rightarrow$ R, at a metal electrode. In each panel, families of parabolae for the free energy (or more precisely, excess electrochemical potential~\cite{bazant2013}) of the oxidized state plus the free electron (O + e$^-$) versus reorganization reaction coordinate that intersect a corresponding parabola for the reduced state (R) are shown on the left, for different electron energies sampled from the Fermi-Dirac distribution shown  on the right. (a) Exchange process at zero overpotential, dominated by electrons near the Fermi level following Marcus kinetics. (b) Reaction-limited current at large positive overpotential, dominated by lower-energy electrons below the Fermi level undergoing barrier-less transitions.          \label{fig:MHC_curves} }
\end{figure}

The disappearance of the inverted region originates from the distribution of electrons in the metal electrode, as shown in Fig.~\ref{fig:MHC_curves}. When a positive free-energy barrier is formed in the inverted region in response to the large overpotential, electrons below the Fermi level ($\mu_e$) with roughly unity Fermi factor follow a lower-energy parabola that enables a barrier-less transfer, which dominates the overall reduction rate and leads to a constant, non-zero limiting current~\cite{oldham1968,hale1968,schmickler1975}. More detailed comparisons between BV and MHC kinetics can be found in Appleby and Zagal~\cite{appleby2011}, Chen and Liu~\cite{chen2014}, and the enlightening review of Henstridge et al.~\cite{henstridge2012}.

Evidence is mounting that MHC kinetics are essential for the understanding and engineering of important electrochemical interfaces.   The MHC model has been extensively used in the microscopic analysis of electron transfer at SAMs~\cite{chidsey1991,henstridge2012} and electrochemical molecular
junctions~\cite{migliore2011}.  It could also be important for nano-electrochemical systems working at large overpotentials, such as resistive-switching memory ~\cite{waser2009} or integrated circuits with ultrathin gate dielectrics, where the BV model predicts unrealistically large reaction rates~\cite{muralidhar2014}. Recent Tafel analysis of Li-ion battery porous electrodes consisting of carbon-coated LiFePO$_4$ particles has further verified MHC kinetics for electron transfer at the carbon-LiFePO$_4$ (solid-solid) interface~\cite{bai2014}, contrary to all  existing battery models, which assume BV kinetics.  

One possible reason MHC kinetics have been overlooked is the complexity of the rate expression Eq.~\ref{eqn:MHC_full} as an improper integral that cannot be evaluated in terms of elementary functions, like the BV equation. In order to avoid numerical quadrature, there have been several attempts to derive simpler analytical approximations. Oldham and Myland~\cite{oldham2011} recently obtained an exact solution involving sums of a function that is a product of an exponential function and a complementary error function, which leads to some convenient alternatives for limited ranges of the parameters. Migliore and Nitzan derived another series solution by an expansion of the Fermi function~\cite{ migliore2011}, which is mathematically equivalent to Oldham's solution~\cite{migliore2012}. As with any series expansion, however, accuracy is lost upon truncation, and the approximations are not uniformly valid across the range of possible reorganization energies and overpotentials.

In this paper, we derive a simple formula by asymptotic matching that accurately approximates the MHC integral over the entire realistic parameter range. In the following sections, we first perform asymptotic analysis of Eq.~\ref{eqn:MHC_full} for positive (oxidation) and negative (reduction) overpotentials, then unify both cases by asymptotic matching in a closed-form approximation, and finally demonstrate the accuracy of our formula compared to numerical quadrature and the recent series solutions.  Complete asymptotic series are derived in the appendices for large and small reorganization energies, but only the leading-order terms are used in the main text to obtain our uniformly valid formula.

% -----------------------------------------------------------------------------------------
% New Section
% -----------------------------------------------------------------------------------------

\section{ Oxidation Rate for Positive Overpotentials }

Without loss of generality, we neglect the prefactor $A$ and begin by restricting $\eta>0$  for the oxidation rate. Equation.~\ref{eqn:MHC_full} can then be rewritten as,
\begin{equation}
\label{eqn:MHC}
k(\lambda,\eta) = \int^{+\infty}_{-\infty} g(x;\lambda, \eta) f(x)  dx,
\end{equation}
where the original integrand is separated to a Gaussian function $g$ and the Fermi distribution $f$,
\begin{equation}
\begin{aligned}
g(x;\lambda, \eta) &= \exp \left( - \frac{(x - \lambda + \eta)^2}{4 \lambda} \right), \\
f(x) &= \frac{1}{1 + \exp(x)}.
\end{aligned}
\label{eqn:f_g_def}
\end{equation}
For mathematical convenience, all quantities starting from Eq.~\ref{eqn:MHC} will be dimensionless: $x$ and $\lambda$ are scaled to $k_BT$ and $\eta$  to $k_BT/e$.

\subsection{ Small reorganization energies, $\lambda \ll 1$}

When $\lambda \ll 1$, the Gaussian function $g$ has a narrow peak at $x = \lambda - \eta$. We will apply the Laplace method~\cite{cheng2006,bender_orzag_book}, where we expand the function $g$ around the point $x = \lambda - \eta$ by Taylor expansion, and then integrate all the terms separately. Derivations and the full series solution can be found in Appendix~\ref{sec:small_lambda}. Here, we use the leading asymptotic term of the integral,
\begin{equation}
k(\lambda,\eta) \approx \frac{2\sqrt{\pi \lambda}}{1 + \exp(\lambda - \eta)},
\label{eqn:small_lambda}
\end{equation}
as our asymptotic approximation for cases of small $\lambda$.

\subsection{Large reorganization energies, $\lambda \gg 1$}

For an outer-sphere reaction, $\lambda$ is usually larger than $1$, and the series solution given in Eq.~\ref{eqn:Taylor_Series} may converge slowly.  A more accurate approximation for the integral in Eq.~\ref{eqn:MHC} in this limit is based on the observation,
\begin{equation}
\lim_{a \rightarrow + \infty} \frac{1}{1 + \exp(ax)} = 1 - H(x),
\end{equation}
where $H(x)$ is the Heaviside step function defined to be $H(x) = 0$ for $x < 0$ and, $H(x) = \frac{1}{2}$ for $x = 0$ and $H(x) = 1$ elsewhere.
This  corresponds to the zero temperature limit of the Fermi-Dirac distribution, which enables an accurate approximation to the original integral~\cite{hale1968},
\begin{equation}
k(\lambda,\eta) \approx  \int^{+\infty}_{-\infty} g(x;\lambda, \eta) \left( 1 - H(x) \right)  dx =  \sqrt{\pi \lambda}  \erfc \left( \frac{\lambda - \eta}{2\sqrt{\lambda}}\right),
\label{eqn:large_lambda}
\end{equation}
where $\erfc(\cdot)$ is the complementary error function. The derivation of the correction series to this approximation is available in Appendix~\ref{sec:large_lambda}.

% -----------------------------------------------------------------------------------------
% New Section
% -----------------------------------------------------------------------------------------

\section{ Oxidation rate for negative overpotentials }

Combining the de Donder relation and reciprocity relations for MHC kinetics~\cite{oldham2011}, we obtain a symmetry condition
\begin{equation}
\label{eqn:redundancy}
\frac{k(\lambda,\eta)}{k(\lambda, -\eta)} = \exp(\eta),
\end{equation}
which directly yields the leading-order approximation for $\eta <0$.
When $\lambda \ll 1$, by using Eq.~\ref{eqn:small_lambda} and Eq.~\ref{eqn:redundancy}, we have,
\begin{equation}
k(\lambda,\eta) = \exp(\eta) k(\lambda,-\eta) 
\approx \frac{2\sqrt{\pi \lambda}\exp(\eta)}{1 + \exp(\lambda + \eta)}.
\end{equation}
And for the case of $\lambda \gg 1$, by using Eq.~\ref{eqn:large_lambda} and Eq.~\ref{eqn:redundancy}, we obtain
\begin{equation}
k(\lambda,\eta) \approx \sqrt{\pi \lambda} \exp(\eta) \erfc \left( \frac{\lambda + \eta}{2\sqrt{\lambda}} \right).
\end{equation}

We thus obtain asymptotic approximations of the integral \ref{eqn:MHC} for all $\eta$, in the limit $\lambda \ll 1$,
\begin{equation}
\label{eqn:lambda_small_approx}
k(\lambda,\eta)  \approx \left\{ \begin{aligned}
	& \frac{2\sqrt{\pi \lambda}}{1 + \exp(\lambda - \eta)} & & \text{\indent for } \eta \geq 0 \text{ and } \lambda \ll 1,  \\
	& \frac{2\sqrt{\pi \lambda}\exp(\eta)}{1 + \exp(\lambda + \eta)} & &\text{\indent for } \eta < 0 \text{ and } \lambda \ll 1,
             \end{aligned}
	\right.
\end{equation}
and the limit $\lambda \gg 1$,
\begin{equation}
\label{eqn:lambda_big_approx}
k(\lambda,\eta)  \approx \left\{ \begin{aligned}
	& \sqrt{\pi \lambda}  \erfc \left( \frac{\lambda - \eta}{2\sqrt{\lambda}}\right) & & \text{\indent for } \eta \geq 0 \text{ and }  \lambda \gg 1,  \\
	& \sqrt{\pi \lambda} \exp(\eta) \erfc \left( \frac{\lambda + \eta}{2\sqrt{\lambda}} \right) & &\text{\indent for } \eta < 0 \text{ and }   \lambda \gg 1. \\
             \end{aligned}
	\right.
\end{equation}

% -----------------------------------------------------------------------------------------
% New Section
% -----------------------------------------------------------------------------------------

\section{ Uniformly Valid Approximation }

In order to get a closed form expression valid for all $\eta$, we multiply the $\eta\geq 0$ approximation by a function $M(\eta)$ that interpolates between the asymptotic limits, $M(\eta)\to 1$ for $\eta \to \infty$ and $M(\eta)\sim e^\eta$ for $\eta\to-\infty$. In order to make the expression differentiable, we also introduce a function $N(\eta)$ to continuously approximate the absolute value function,
\begin{equation}
k(\lambda,\eta)  \approx  \sqrt{\pi \lambda} M(\eta)  \erfc\left(\frac{\lambda - N(\eta)}{2\sqrt{\lambda}}\right). 
\end{equation}
Although it is possible to also construct a uniformly valid approximation for all $\lambda$ in a similar way, we consider only the $\lambda \gg 1$ approximation, which turns out to be accurate even down to $\lambda \approx 0.1$ and covers the physically relevant range for outer sphere reactions.  Below such small values of the reorganization energy, the barrier to charge transfer is too small to justify the use of transition state theory, and MHC kinetics break down.    

For smooth $M(\eta)$ and $N(\eta)$, the uniformly valid approximation removes the discontinuous derivative at $\eta=0$ that would arise by naively patching the two asymptotic approximations for $\eta >0$ and $\eta <0$. The de Donder relation can also be satisfied exactly if we require $M(\eta)=e^{\eta}M(-\eta)$.  These properties are satisfied by the following simple choices for the interpolating functions
\begin{equation}
\begin{aligned}
M(\eta) &= \frac{1}{1 + \exp(-\eta)}, \\
N(\eta) &= \sqrt{a + \eta^2}, 
\end{aligned}
\end{equation}
where $a$ is an arbitrary constant, yielding the uniformly valid approximation
\begin{equation}
\label{eqn:asymptotic_approx}
k(\lambda,\eta)  \approx \frac{ \sqrt{\pi \lambda}}{1 + \exp(-\eta)}  \erfc\left(\frac{\lambda - \sqrt{a + \eta^2}}{2\sqrt{\lambda}}\right).
\end{equation}

\begin{figure} [h]
\begin{center}
\begin{tabular}{ c c}
\includegraphics[width=0.5 \columnwidth]{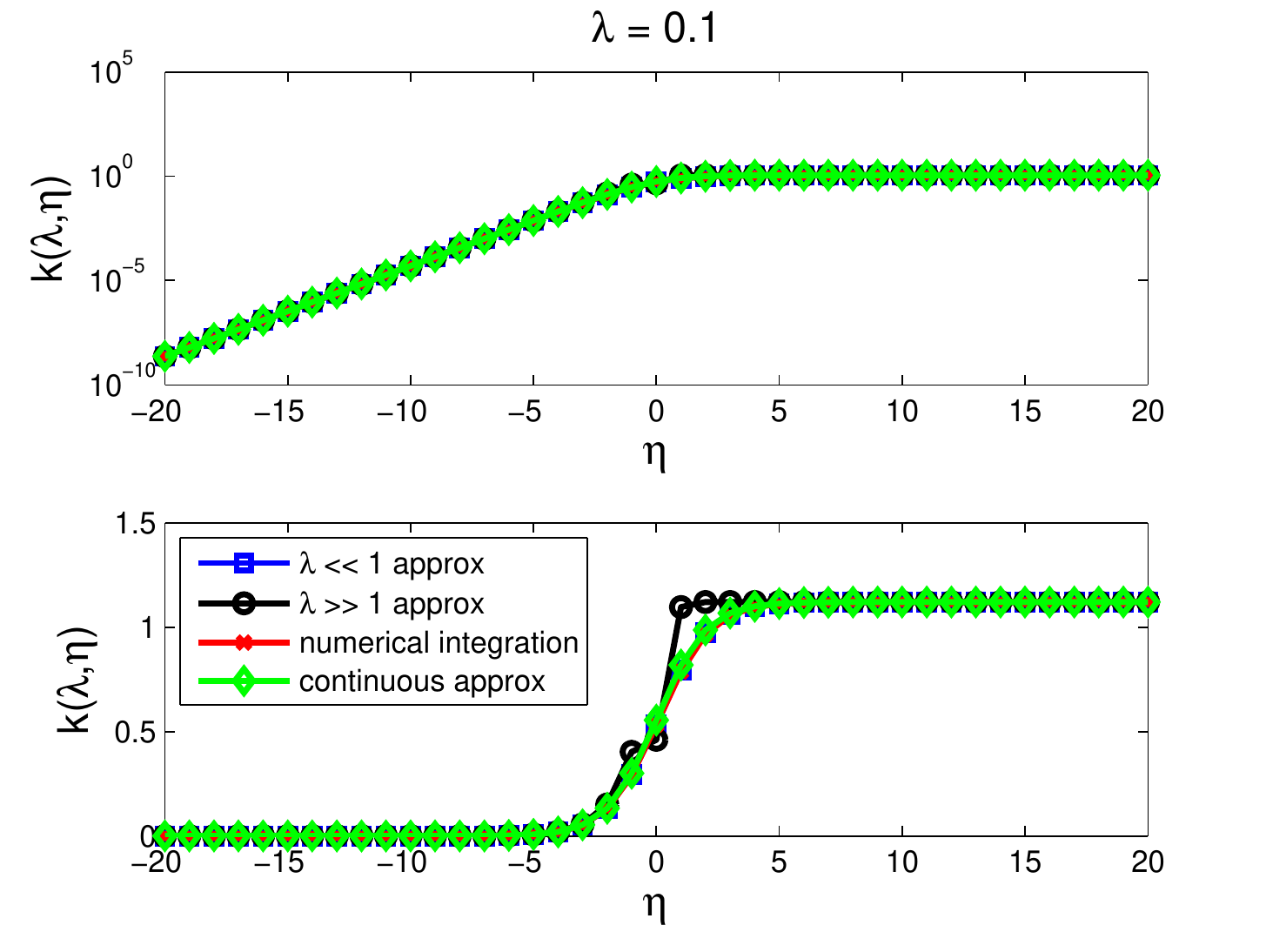} & \includegraphics[width=0.5 \columnwidth]{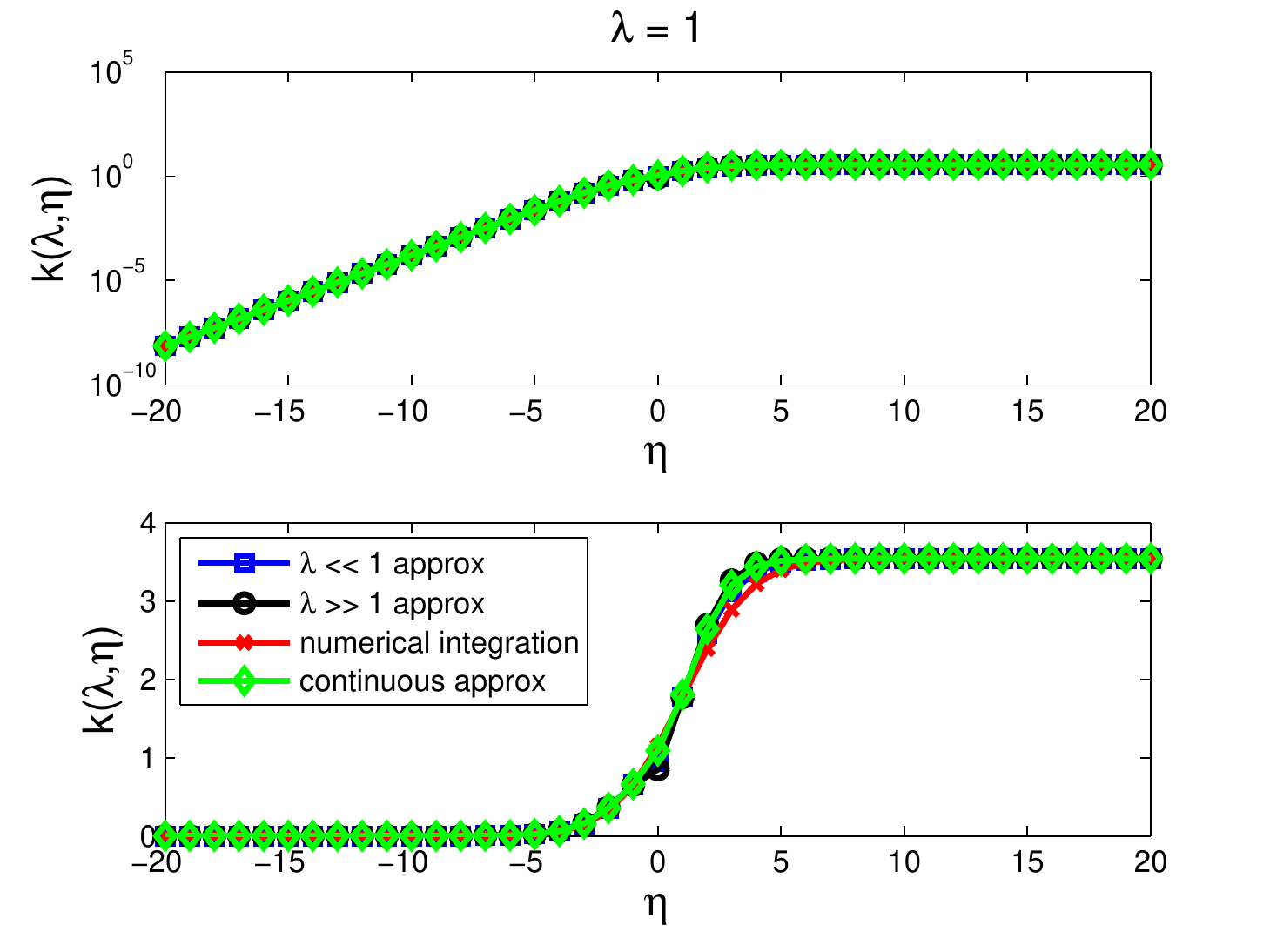} \\
\includegraphics[width=0.5 \columnwidth]{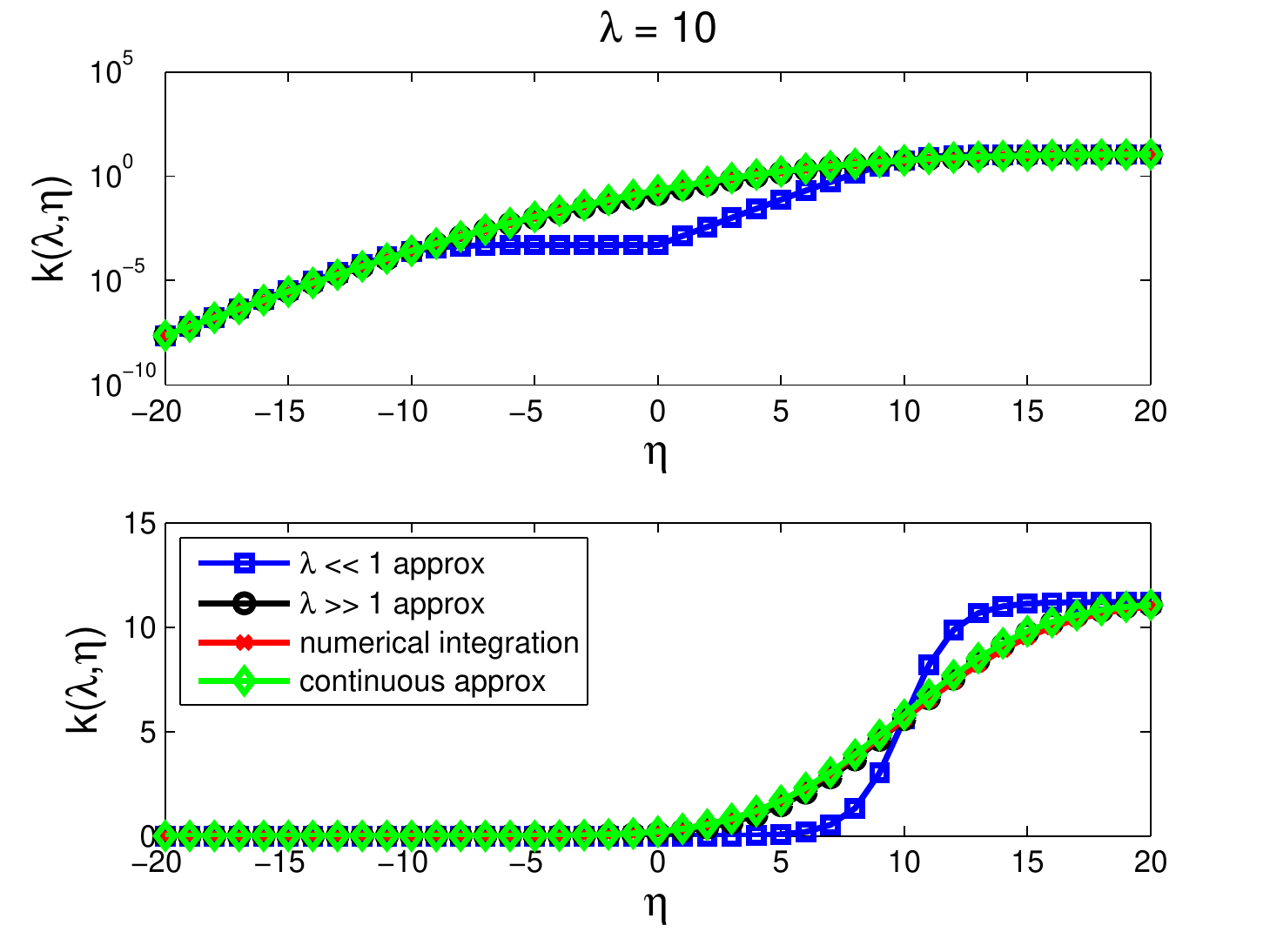} & \includegraphics[width=0.5 \columnwidth]{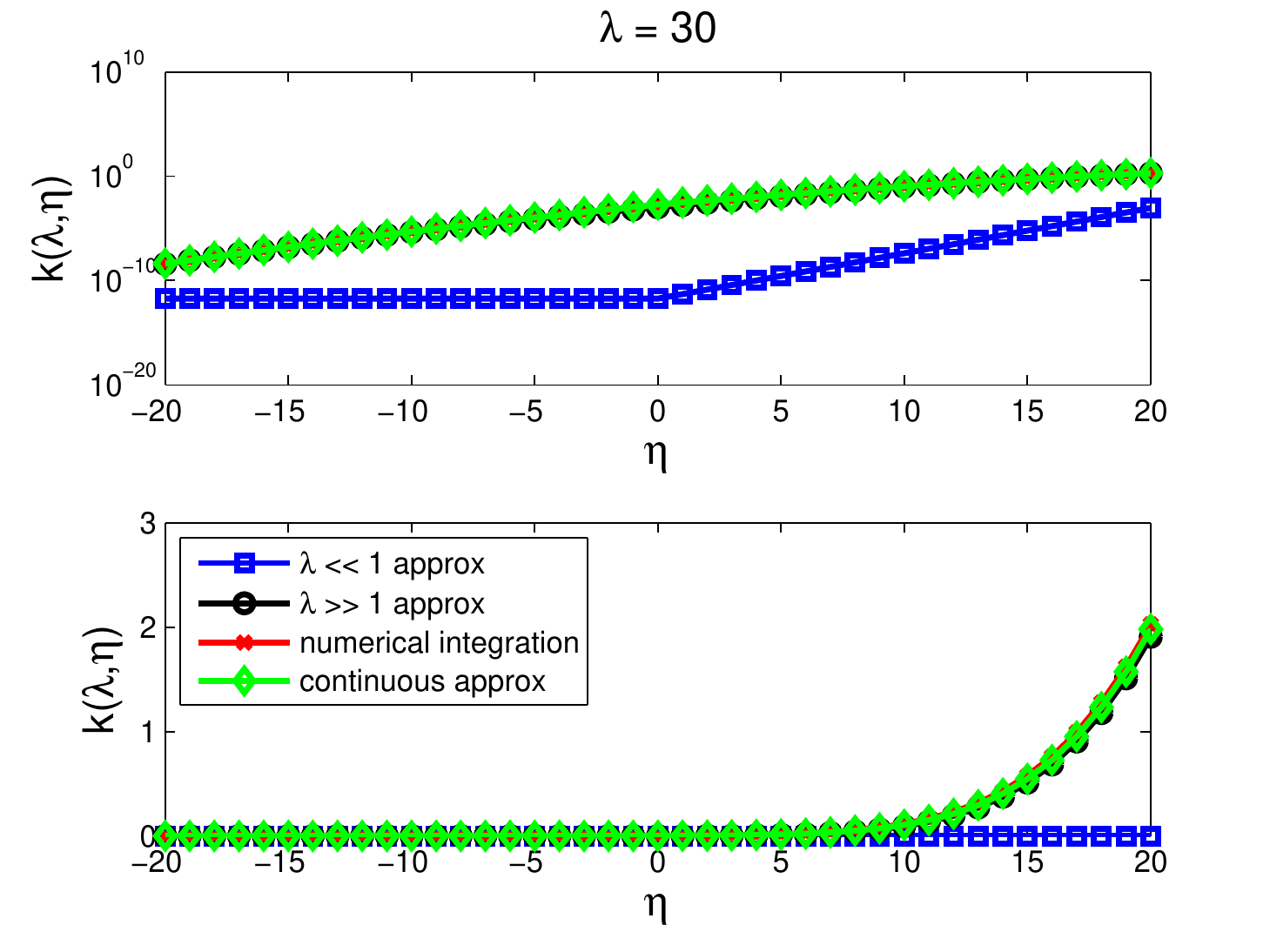}
\end{tabular}
\end{center}
\caption{Numerical evaluations of reaction rates $k(\lambda,\eta)$ according to three asymptotic approximations Eq.~\ref{eqn:lambda_small_approx} (blue square), Eq.~\ref{eqn:lambda_big_approx} (black circle) and Eq.~\ref{eqn:asymptotic_approx} (green diamond), together with the direct numerical quadrature of the MHC integral~\ref{eqn:MHC} (red cross) for $\lambda = 0.1$, $1$, $10$ and $30$ and $|\eta|<20$. Each comparison is shown in both log scale (top) and linear scale (bottom).}
\label{fig:ErrorConvergence1}
\end{figure}

A comparison between different approximations (small $\lambda$ limit, large $\lambda$ limit, and uniform approximation) and the direct numerical integration of MHC for various $\lambda$ values are shown in Figure~\ref{fig:ErrorConvergence1}. Remarkably, we find that Eq.~\ref{eqn:asymptotic_approx} with $a = 1 + \sqrt{\lambda}$ provides very accurate approximation to the MHC integral (Eq.~\ref{eqn:MHC}) across the full range of physical parameter values. The numerical results almost overlap  everywhere, as shown in Fig.~\ref{fig:ErrorConvergence1}.

\begin{figure} [h]
\begin{center}
(a)\includegraphics[width=3in]{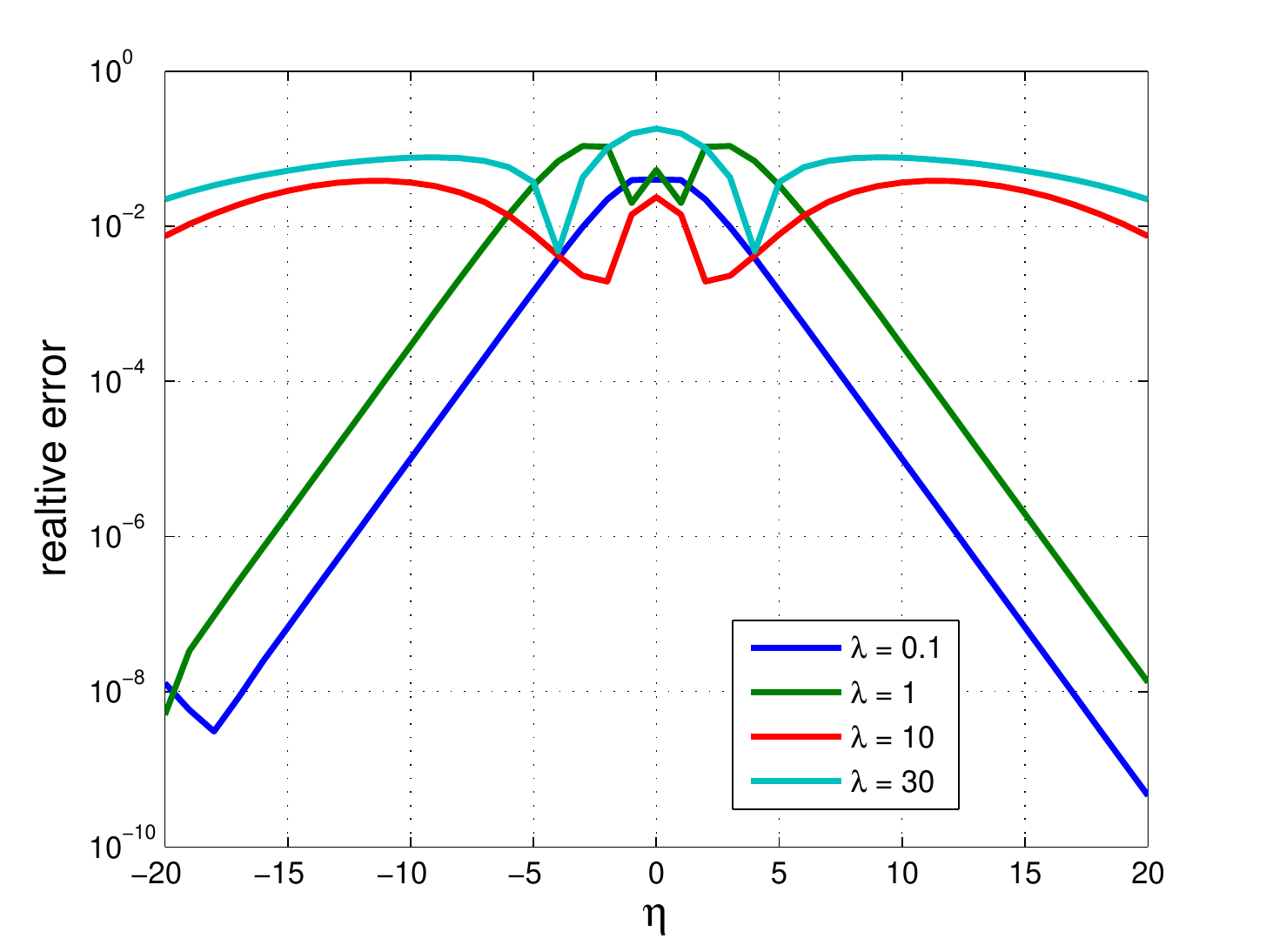}\\
(b)\includegraphics[width=3in]{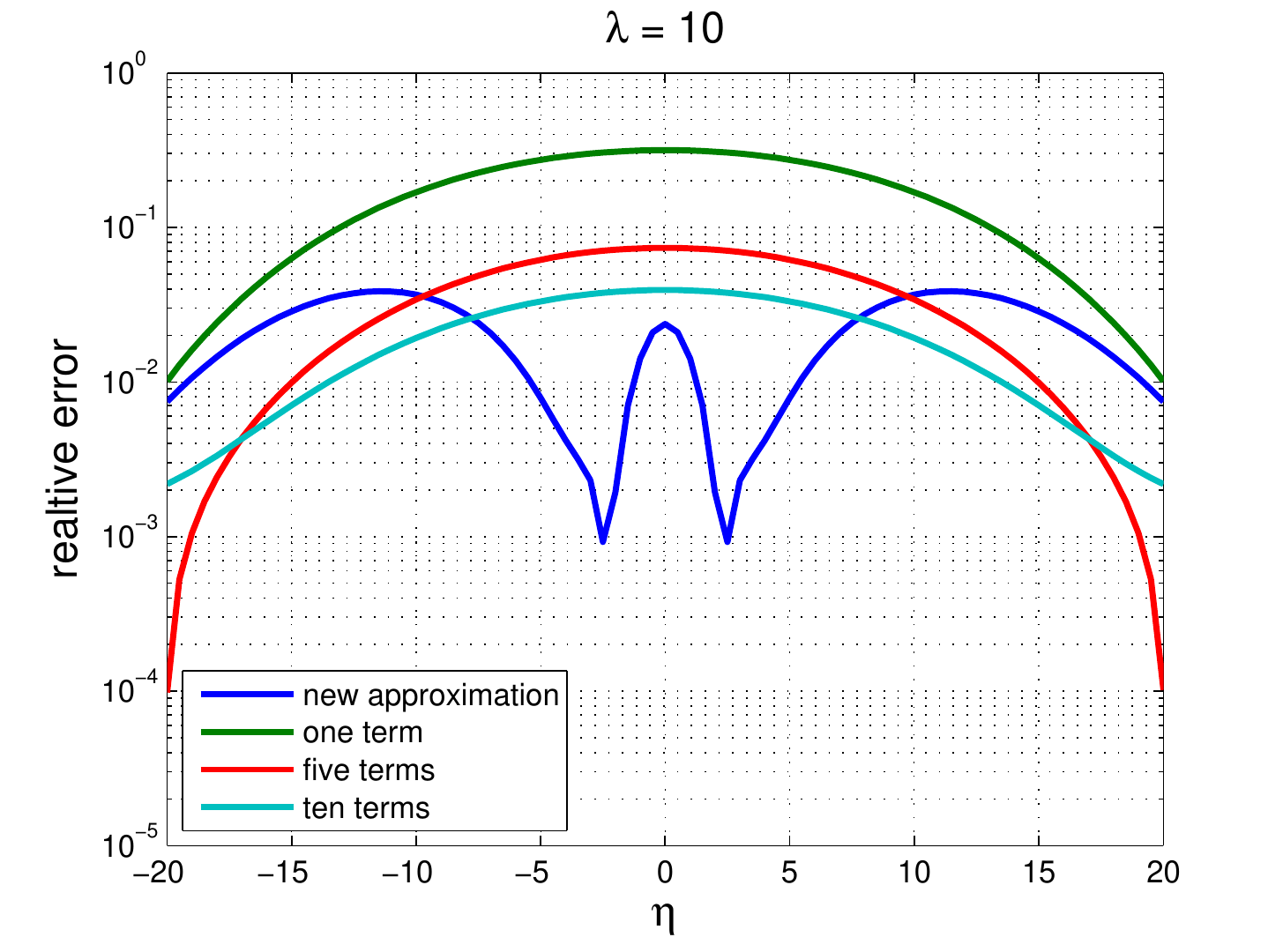} 
\end{center}
\caption{(a) Relative error of our simple formula Eq.~\ref{eqn:asymptotic_approx} compared to numerical quadrature of the MHC integral~\ref{eqn:MHC} for $\lambda = 0.1$, $1$, $10$ and $30$ and  $|\eta|<20$. (b) Relative error of our formula for $\lambda=10$ compared with the series approximation of Oldham and Myland \cite{oldham2011} with $1$, $5$ and $10$ terms.}
\label{fig:ErrorConvergence2}
\end{figure}

Numerical evaluations of the relative errors of our simple formula~\ref{eqn:asymptotic_approx} under different choices of $\lambda$ are shown in Fig.~\ref{fig:ErrorConvergence2}, including a comparison with the series solution by Oldham and Myland~\cite{oldham2011} for $\lambda = 10$. It is clearly seen that our approximation exhibits $<10\%$ relative error even in the most extreme cases. For more relevant cases for outer sphere reactions (e.g. $\lambda \approx 10$)~\cite{bai2014,chidsey1991}, the relative error is less than $5\%$ for small overpotentials and vanishingly small at large positive or negative overpotentials.

Finally, we arrive at our main result. By subtracting the oxidation rate from the reduction rate, $I(\lambda,\eta) = k(-\eta,\lambda) - k(\eta,\lambda)$,  we obtain a simple, accurate, formula for the net reduction current (up to a constant pre-factor):
\begin{equation}
I(\lambda,\eta) \approx \sqrt{\pi \lambda} \tanh\left(\frac{\eta}{2}\right) \erfc\left(\frac{\lambda - \sqrt{1 + \sqrt{\lambda} + \eta^2}}{2\sqrt{\lambda}}\right).    \label{eq:formula}
\end{equation}
This expression is almost as simple and efficient to evaluate as the BV equation, while accurately approximating the MHC integral over the entire physical parameter range. For example, on a dual-core processor using Python with Scipy, the evaluation of  Eq. \ref{eq:formula} is only about four times slower than that of the BV equation, but about 1500 times faster than an efficient numerical quadrature of the MHC integral using a subroutine from the Fortran QUADPACK library (with $\lambda = 10$).   

\begin{figure}
\centering
\includegraphics[width=3in]{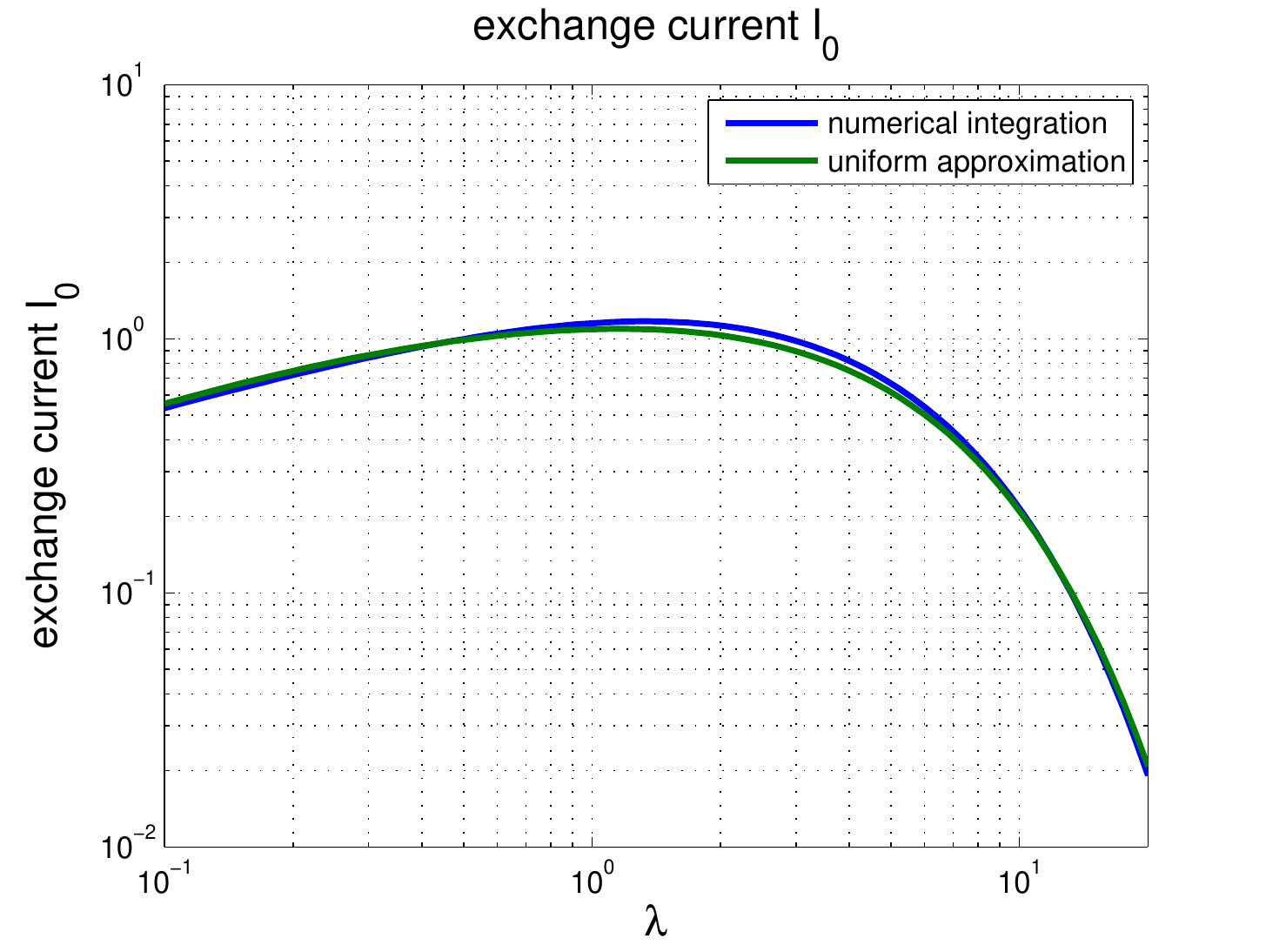}
\caption{   Dimensionless exchange current $k(\lambda,\eta=0)$ versus reorganization energy $\lambda$ for numerical quadrature of the MHC integral compared to the uniformly valid approximation, Eq. \ref{eq:I0}, showing maximum 5\% error when $0.1 \leq \lambda \leq 20$.  \label{fig:ex} }
\end{figure}

From Eq. ~\ref{eq:formula}, the exchange current (up to the same constant) is the forward or backward rate in equilibrium,
\begin{equation}
I_0(\lambda) = k(\lambda,0) \approx \frac{ \sqrt{\pi \lambda} }{2} \erfc\left( \frac{ \lambda - \sqrt{1 + \sqrt{\lambda}} }{2\sqrt{\lambda}}\right),     \label{eq:I0}
\end{equation}
which decays exponentially for large reorganization energies,
\begin{equation}
I_0 \approx \exp\left( - \frac{\lambda}{4} \right), \text{\indent} \lambda \gg 1.
\end{equation}
Although the greatest error in our formula occurs at small over-potentials (Fig. \ref{fig:ErrorConvergence2}),  the accuracy is quite satisfactory even at $\eta=0$ for a wide range of reorganization energies, as shown in Fig. ~\ref{fig:ex}.

% -----------------------------------------------------------------------------------------
% New Section
% -----------------------------------------------------------------------------------------

\section{Conclusion}
In order to facilitate the application of the MHC kinetics in electrochemical engineering, we derive a simple approximation by asymptotic matching that serves as a practical alternative to the BV equation for electrochemical engineering.  Our formula improves upon classical asymptotic approximations~\cite{oldham1968,hale1968,schmickler1975} and recent series expansions~\cite{oldham2011,migliore2011,migliore2012} and provides the first uniformly valid approximation for all reasonable choices of the reorganization energy and overpotential  with less than 5\% error at small overpotentials and vanishing error at large overpotentials. This result could be conveniently used in classical battery models~\cite{newman_book} or new models based on non-equilibrium thermodynamics~\cite{bazant2013} for electrode phase transformations limited by Faradaic reactions~\cite{bai2014}.  Switching from BV to MHC kinetics could have significant implications for the understanding and optimization of electrochemical systems working at high overpotentials.

% -----------------------------------------------------------------------------------------
% New Section
% -----------------------------------------------------------------------------------------

\section*{Acknowledgments}

This work was supported by the National Science Foundation Graduate Research Fellowship under Grant No. 1122374 (Y. Z.) and by the Samsung-MIT Alliance.

% -----------------------------------------------------------------------------------------
% Appendix
% -----------------------------------------------------------------------------------------

\appendix

\section{Small $\lambda$ Limit}
\label{sec:small_lambda}

The Taylor series of the Fermi distribution function $f$ defined in Eq.~\ref{eqn:f_g_def} around $x = \lambda - \eta$ is,
\begin{equation}
f(x) = \sum_{n = 0}^{\infty} \frac{(x - \lambda + \eta)^n}{n !} f^{(n)}(\lambda - \eta).
\end{equation}
If we put this expression back to Eq.~\ref{eqn:MHC}, we get,
\begin{equation}
\begin{aligned}
k(\lambda,\eta) = & 2\sqrt{\pi \lambda} \sum_{n = 0}^{\infty}  \frac{f^{(n)}(\lambda - \eta)}{n !} \\
& \int^{+\infty}_{-\infty} \frac{1}{2\sqrt{\pi \lambda}} (x - \lambda + \eta)^n \exp \left( - \frac{(x - \lambda \pm \eta)^2}{4 \lambda} \right) dx.
\end{aligned}
\end{equation}
For each $n$, the integral is exactly the n-th central moment of a normal distribution with variance $\sigma^2 = 2\lambda$, then the value for such an integration is,
\begin{eqnarray*}
& \int^{+\infty}_{-\infty} \frac{1}{2\sqrt{\pi \lambda}} (x - \lambda + \eta)^n \exp \left(- \frac{(x - \lambda \pm \eta)^2}{4 \lambda} \right) dx \\
& = \left\{ \begin{aligned}
	&1 & & \text{\indent for } n = 0, \\
	&0 & &\text{\indent for $n$ is odd}, \\
	&(2\lambda)^{\frac{n}{2}} (n-1)!! & &\text{\indent for $n>0$ is even}.
             \end{aligned}
	\right.
\end{eqnarray*}
Therefore, the series for $k(\lambda,\eta)$ is,
\begin{equation}
\label{eqn:Taylor_Series}
k(\lambda,\eta) = 2\sqrt{\pi \lambda} \sum_{n = 0}^{\infty}  \frac{\lambda^{n} }{n!} f^{(2n)}(\lambda - \eta).
\end{equation}

\section{Large $\lambda$ Limit}
\label{sec:large_lambda}

For large $\lambda$, we first rewrite Eq.~ \ref{eqn:MHC} as,
\begin{equation}
\begin{aligned}
\label{eqn:Heaviside_appro}
k(\lambda,\eta) &= \int^{+\infty}_{-\infty} g(x;\lambda, \eta) \left( 1 - H(x) \right)  dx \\
& + \int^{+\infty}_{-\infty} g(x;\lambda, \eta) \left( f(x) - 1 + H(x) \right) dx.
\end{aligned}
\end{equation}
The first term on the right hand side of Eq.~\ref{eqn:Heaviside_appro} can be exactly solved as shown in Eq.~\ref{eqn:large_lambda}, while the second half can be simplified to,
\begin{equation}
\begin{aligned}
\label{eqn:error_term}
& \int^{+\infty}_{-\infty} g(x;\lambda, \eta) \left(f(x) - 1 + H(x)\right)  dx  \\
& = -2 \exp \left( -\frac{(\lambda - \eta)^2}{4 \lambda} \right) \int^{+\infty}_{0} \exp \left(-\frac{x^2}{4\lambda}\right)  \frac{\sinh \frac{(\lambda - \eta) x}{2 \lambda} }{1 + \exp(x)} dx.
\end{aligned}
\end{equation}
If we define a new function $h$ as,
\begin{equation}
h(x) = \frac{\sinh \frac{(\lambda - \eta) x}{2 \lambda} }{1 + \exp(x)},
\end{equation}
since $h(x=0) = 0$, its Maclaurin series is,
\begin{equation}
h(x) = \sum_{n = 1}^{\infty} \frac{x^n}{n!}h^{(n)}(0).
\end{equation}
We substitute this back to Eq.~\ref{eqn:error_term} and obtain,
\begin{equation}
\begin{aligned}
& \int^{+\infty}_{-\infty}  g(x;\lambda, \eta) \left( f(x) - 1 + H(x) \right)  dx \\
&=-2 \exp\left( -\frac{(\lambda - \eta)^2}{4 \lambda} \right) \sum_{n = 1}^{\infty} \frac{h^{(n)}(0)}{n!}  \int^{+\infty}_{0} x^n \exp(-\frac{x^2}{4\lambda}) dx \\
&= -2 \exp\left( -\frac{(\lambda - \eta)^2}{4 \lambda} \right) \sum_{n = 1}^{\infty} h^{(n)}(0) \frac{2^n}{n!} \lambda^{\frac{n+1}{2}} \Gamma\left(\frac{n+1}{2} \right),
\end{aligned}
\end{equation}
where $\Gamma(\cdot)$ is the gamma function. Thus, the MHC integral in Eq.~\ref{eqn:MHC} can be expanded asymptotically as,
\begin{equation}
\begin{aligned}
\label{eqn:Heaviside_error_expansion}
& k(\lambda,\eta) = \sqrt{\pi \lambda}  \erfc \left( \frac{\lambda - \eta}{2\sqrt{\lambda}}\right) \\
& -2 \exp\left( -\frac{(\lambda - \eta)^2}{4 \lambda} \right) \sum_{n = 1}^{\infty} h^{(n)}(0) \frac{2^n}{n!} \lambda^{\frac{n+1}{2}} \Gamma \left(\frac{n+1}{2}\right).
\end{aligned}
\end{equation}

\renewcommand\refname{Reference}
\bibliographystyle{plain}
\bibliography{elec44}

\end{document}